\documentclass[journal]{IEEEtran}

\ifCLASSINFOpdf
\else
\fi

\usepackage{graphicx}
\usepackage{amsmath}
\usepackage{xcolor}
\usepackage{tabularx}
\usepackage{booktabs}
\usepackage{makecell}
\usepackage{array}
\usepackage[table]{xcolor}
\usepackage{hyperref}
\hypersetup{
    colorlinks=true,
    linkcolor=blue,
    citecolor=blue,
    urlcolor=blue}

\begin{document}

\title{Noise-Robust Frequency Estimation via Overlapped Sampling-Intervals Zero-Crossing Fitting}

\author{Phichai Youplao, Nithiroth Pornsuwancharoen, and Yusaku Fujii~% <-this % stops a space

\thanks{This is the accepted version of the article published in IEEE Transactions on Instrumentation and Measurement. The final version is available at https://doi.org/10.1109/TIM.2026.3693432. © 2026 IEEE.}
}

\maketitle

\begin{abstract}
The trade-off between noise averaging and temporal resolution fundamentally limits conventional zero-crossing frequency estimators under dynamic and noisy conditions. This paper presents an overlapped sampling-intervals zero-crossing fitting method (OS-ZFM), which introduces a structured overlapping regression framework that decouples noise averaging from temporal update rate. The method adopts a deterministic closed-form formulation, enabling a unified bias-variance analysis to characterize the statistical behavior of the estimator and clarify the role of structured data reuse. Numerical results under intensity and background noise at a signal-to-noise ratio (SNR) of 10 dB show that OS-ZFM reduces median estimation error by more than 60\% compared to conventional zero-crossing fitting methods at the same temporal resolution. It further achieves up to 90\% reduction relative to basic zero-crossing detection and consistently yields lower estimation errors than Hilbert-transform-based estimators. Experimental validation utilizing impact-induced transient motion measured by laser Doppler interferometry demonstrates that OS-ZFM reconstructs smooth and physically consistent trajectories with improved temporal fidelity. Owing to its low computational complexity and deterministic formulation, the proposed method enables accurate real-time frequency and acceleration tracking in resource-constrained measurement systems.

\end{abstract}


\IEEEpeerreviewmaketitle

% ==================================================================
\section{Introduction}

\IEEEPARstart{A}{ccurate} frequency estimation is a fundamental task in instrumentation and measurement and underpins a wide range of sensing applications, including vibration analysis \cite{Li2020, Li2024}, precision displacement sensing \cite{Dre2013, Shen2021}, dynamic force and acceleration measurement \cite{Liu2025, Youplao2024}, optical interferometric sensing \cite{Hor2019, Youplao2021}, and laser Doppler-based motion measurement \cite{Xiang2025, Yang2026}. Recent advances in frequency estimation, such as adaptive spectral analysis and wavelet-based methods \cite{Zhao2023, Sun2025}, time--frequency reassignment techniques \cite{Tazen2023}, state-space modeling \cite{Lee2024}, and statistically optimal estimators \cite{Aharon2024}, have demonstrated excellent performance under well-controlled conditions when sufficient computational resources and accurate signal models are available. In practical measurement systems, however, frequency estimation must satisfy performance requirements beyond estimation accuracy. Embedded and real-time instrumentation, particularly laser Doppler interferometers used for dynamic motion and force measurements, demands low computational complexity, minimal latency, robustness to amplitude fluctuations and background noise, and transparent estimation mechanisms. Moreover, measurement-oriented applications require predictable noise behavior and interpretability to support uncertainty analysis and metrological traceability—requirements that are not always readily met by highly sophisticated or model-intensive estimation techniques.

Although modern frequency estimation approaches exhibit strong theoretical performance, many—including dense spectral processing, iterative optimization, phase-based estimation, and model-intensive regression—may degrade under nonstationary conditions, amplitude modulation, and hardware-induced distortions commonly encountered in practical interferometric measurements \cite{Liu2022, Luo2025}. In addition, their computational burden and parameter-tuning requirements can hinder real-time implementation in resource-constrained systems \cite{Be2024}. Consequently, time-domain frequency estimation methods that operate directly on digitized waveforms, rely on minimal assumptions, and exhibit transparent noise behavior remain of strong practical interest \cite{Kun2025, Tan2024}.

Among such approaches, zerocrossing-based frequency estimation—where the frequency is inferred from the estimated time instants at which the signal crosses zero amplitude—has attracted sustained attention due to its simplicity and low computational cost \cite{Serbes2020, Amin2026}. Conventional zero-crossing methods (ZCM), which rely on only two boundary crossings per interval, are highly sensitive to noise-induced timing jitter, leading to large estimation variance under low signal-to-noise ratio (SNR) or rapidly varying frequency dynamics. The zero-crossing fitting method (ZFM) mitigates this issue by exploiting all zero-crossing points within an observation window—the set of consecutive samples used for local frequency estimation—via closed-form least-squares regression \cite{Fujii2009}, thereby improving noise robustness while maintaining computational efficiency. Nevertheless, ZFM still faces a fundamental trade-off: longer windows improve noise averaging but introduce temporal smoothing and tracking bias, whereas shorter windows enhance responsiveness at the cost of increased variance.

% -------------------------------------------------------------
\begin{figure*}[!t]
\centering
\includegraphics[width=6.8in]{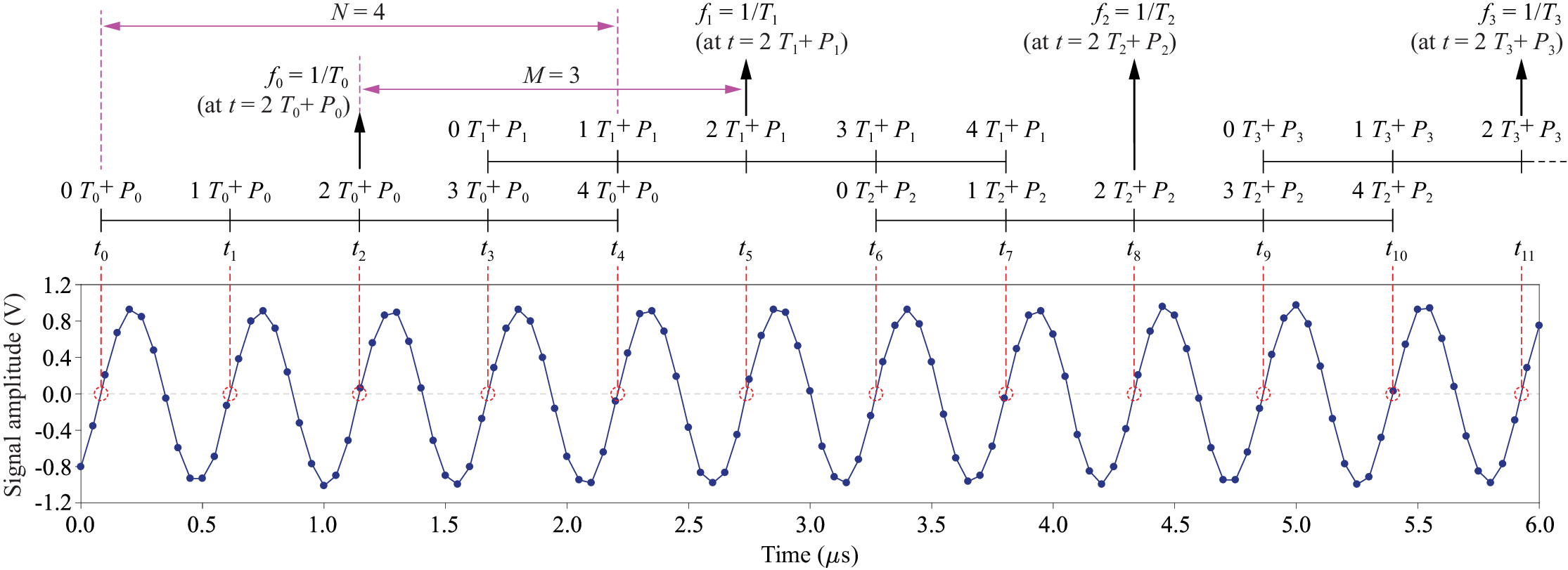}
\caption{Algorithm of the OS-ZFM, where the observation window periods are defined as $N=\text{4}$ and $M=\text{3}$.}
\label{algorithm}
\end{figure*}
% ------------------------------------------------------------

Despite extensive studies on zero-crossing-based estimators, this trade-off has not been structurally resolved within the regression-based framework. Existing methods implicitly select a fixed compromise through window length selection, without decoupling temporal resolution from noise averaging.

This work introduces the Overlapped Sampling-intervals Zero-crossing Fitting Method (OS-ZFM) for frequency estimation in interferometric measurements. By utilizing an overlapping regression structure, OS-ZFM decouples the update interval from the observation window, enhancing temporal responsiveness while maintaining noise robustness comparable to ZFM. The method is analyzed via a bias-variance trade-off and evaluated against representative frequency- and phase-based methods, including ZCM, ZFM \cite{Fujii2009}, Hilbert phase-based motion estimation (HPME) \cite{Kun2025}, and the multipoint least-squares (MPLS) method \cite{Tan2024}. Validation is performed through numerical simulations under intensity and background noise, alongside laser Doppler interferometer experiments under transient impact excitation.

The remainder of this paper is organized as follows. Section~II describes the OS-ZFM formulation. Section~III presents the modeled interferometric signals used for numerical evaluation. Section~IV reports simulation results, Section~V provides experimental validation, Section~VI discusses practical implications, and Section~VII concludes the paper.

% ==========================================================
\section{OS-ZFM Formulation}

\subsection{Overlapped Zero-Crossing Fitting Framework}

Fig.~\ref{algorithm} depicts the OS-ZFM algorithm. The harmonic model is segmented into intervals of $N=\text{4}$ periods, with a step size of $M=\text{3}$ periods between adjacent interval centers. Each segment is characterized by its period $T_j$ and phase $\text{2}\pi P_j/T_j$. The least-squares method minimizes the time differences between the zero-crossing points of the model function and those of the sampled signal. This is achieved by independently varying a collection of multiple periods ${T_j}$ and ${P_j}$. Consequently, for the $j$-th interval containing $N+\text{1}$ zero-crossings $t_k$, the frequency $f_j=\text{1}/T_j$ is obtained by minimizing
\begin{equation}
C_j(T_j, P_j) = \sum_{i=0}^{N} \left[t_{i+jM} - (iT_j + P_j)\right]^2,
\label{eq01}
\end{equation}
which leads to the closed-form solution
\begin{equation}
f_j = \frac{4\sum_{i=0}^{N} i^2 - N(N+1)}{4\sum_{i=0}^{N} it_{i+jM} - 2N\sum_{i=0}^{N} t_{i+jM}}.
\label{eq02}
\end{equation}

The zero-crossing time $t_k$ is estimated by linear interpolation between adjacent samples surrounding the negative-to-positive zero-crossing points. The numerator in~\eqref{eq02} is a strictly positive constant for $N > 1$, while the denominator remains well-conditioned due to the monotonic nature of the detected zero-crossing timestamps $t_i$. For sinusoidal signals, the maximum interpolation-induced timing error is approximated as
\begin{equation}
\Delta t_{\max} \approx \frac{\omega^2 T_s^3}{24},
\end{equation}
where $\omega=2\pi f$ and $T_s$ is the sampling period. Under the considered conditions (e.g., $f \approx 1.5$ MHz and $T_s = 50$ ns), this yields an error on the order of tens of picoseconds, which is negligible compared to noise-induced timing jitter.

In OS-ZFM, the observation window length \(N\) governs noise averaging, while the step size \(M\) determines the temporal spacing of successive estimates. When \(M < N\), overlapping intervals are formed, enabling improved temporal sampling without sacrificing noise robustness.

For performance evaluation, conventional ZCM, ZFM \cite{Fujii2009}, HPME \cite{Kun2025}, and MPLS \cite{Tan2024} are used as representative reference methods under identical simulation and noise conditions.

% ----------------------------------------------------------------
\subsection{Distinction from Conventional Sliding-Window Estimation}

Sliding or overlapping windows are commonly employed to increase the output rate of an estimator. In many implementations, however, overlap is applied as a post-processing strategy without modifying the underlying estimator structure.

In conventional zero-crossing fitting, frequency estimation is performed over a block of \(N\) signal periods, where the observation window simultaneously determines both the noise-averaging length and the temporal interval between successive estimates. As a result, increasing \(N\) improves noise robustness but reduces the update rate. The proposed OS-ZFM modifies this structure by introducing partially overlapping regression windows. Specifically, each frequency estimate is obtained by solving a least-squares regression problem over \(N\) consecutive signal periods defined by detected zero-crossing samples. Instead of advancing the observation window by \(N\) periods, the starting index is shifted by \(M < N\), resulting in overlapping observation intervals. This formulation yields a sequence of regression problems, each defined over a fixed-length observation window but with shared samples between adjacent windows. Consequently, the observation length governing noise averaging remains determined by \(N\), while the update interval is controlled by \(M\).

In contrast to conventional sliding-window implementations, the proposed approach incorporates overlap directly into the estimator construction. Each estimate is therefore the solution of a well-defined regression problem rather than an interpolated or smoothed output, representing a structural distinction from conventional approaches. This structurally distinct estimator effectively decouples noise averaging from the temporal update rate, enabling improved temporal resolution while preserving noise robustness.

% --------------------------------------------------------
\subsection{Bias--Variance Interpretation}

The statistical behavior of OS-ZFM can be interpreted by examining how overlapping estimation windows modify the temporal structure of the estimates while preserving the underlying estimator.

Assuming additive noise on the interferometric signal, the detected zero-crossing times are modeled as
\begin{equation}
t_k = kT + \epsilon_k, \qquad k = 0,1,\ldots,
\label{eq:bv_model}
\end{equation}
where $T$ is the true period and $\epsilon_k$ denotes zero-mean timing jitter with variance $\sigma_\epsilon^2$. The independence assumption is adopted for analytical tractability. In practice, bandlimited noise and finite receiver bandwidth may introduce weak correlations between adjacent zero-crossing timing errors. Such correlation does not introduce bias under the zero-mean assumption, but may slightly modify the variance through the covariance structure of the timing errors.

Under this model, the period estimation problem in~\eqref{eq01} corresponds to a linear least-squares regression between the integer index $i$ and the measured zero-crossing times $t_{i+jM}$. Since the measurement noise has zero mean, the resulting least-squares estimator remains unbiased for the signal period $T$. The variance of the estimator is expressed as
\begin{equation}
\mathrm{Var}(\hat{T}) = \frac{\sigma_{\epsilon}^{2}}{\sum_{i=0}^{N} (i-\bar{i})^{2}},
\label{eq:ls_variance}
\end{equation}
where $\bar{i}$ denotes the mean value of the index set $\{0,1,\dots,N\}$.

For equally spaced indices \(i=0,1,\dots,N\), the denominator in~\eqref{eq:ls_variance} admits the closed-form expression
\begin{equation}
\sum_{i=0}^{N}(i-\bar{i})^2=\frac{N(N+1)(N+2)}{12}.
\end{equation}

Substituting this result into~\eqref{eq:ls_variance} shows that the variance of the period estimator decreases asymptotically as
\begin{equation}
\mathrm{Var}(\hat{T}) \approx \frac{12\sigma_\epsilon^2}{N^3}, \qquad (N \gg 1),
\end{equation}
which scales as $O(N^{-3})$ for large $N$. This demonstrates that increasing the observation window length improves noise averaging and reduces estimation variance.

Importantly, the variance depends only on the observation length $N$ and the jitter variance, and is independent of the temporal spacing between estimates. Therefore, the overlapping mechanism in OS-ZFM does not alter the bias or variance of individual estimates, as each estimate is obtained from a regression over the same sample size. Instead, the overlap modifies only the temporal spacing between estimates. By shifting the regression window by $M < N$, the update interval is reduced from $NT$ to $MT$, thereby increasing the temporal sampling density of the estimated frequency trajectory.

Since adjacent windows share samples, successive estimates are partially correlated. As a first-order approximation, the correlation coefficient scales with the fraction of shared samples, i.e., $\rho \propto (N-M)/N$. Thus, while $M \approx N$ yields nearly independent estimates, a smaller $M$ increases correlation. This affects the temporal interpretation of the estimate sequence while preserving the bias and variance of individual estimates.

The classical Cramér-Rao lower bound (CRLB) assumes full access to the continuous waveform, which is inconsistent with zero-crossing estimators operating on discrete timing information. This limitation arises because zero-crossing estimators operate on quantized timing events rather than continuous waveform samples, thereby violating the underlying assumptions required for classical CRLB derivation. Instead, we adopt a theoretical performance baseline derived from the statistical variance of the standard non-overlapping ZFM ($M = N$) with an identical window length. This analytical floor characterizes the performance limit for independent zero-crossing observations where no temporal data reuse is exploited. Given that in interferometric measurements, the target velocity $v(t)$ is directly proportional to the Doppler frequency shift $f_d(t)$ as $v(t) = (\lambda/2)f_d(t)$. Acceleration is subsequently obtained through the temporal differentiation of the velocity trajectory. Specifically, for a non-overlapping baseline, the central-difference approximation is applied to successive velocity estimates separated by $2T_w$. Thus, the baseline estimation variance is formulated as
\begin{equation}
\mathrm{Var}(\hat{a}_{\mathrm{base}})
= \left( \frac{\lambda}{2} \right)^2 \frac{2\sigma_f^2}{(2T_w)^2}
= \frac{\lambda^2 \sigma_f^2}{8T_w^2},
\label{var_a}
\end{equation}
where $\sigma_f^2$ is the variance of a single frequency estimate, $\lambda$ is the signal wavelength, and $T_w$ is the window duration. This benchmark serves as a rigorous reference for assessing the relative performance gains of OS-ZFM.

Based on this analytical baseline, the fundamental distinction between the proposed structural overlap in OS-ZFM and two superficially similar data-densification approaches, output upsampling and post-filtering, can be characterized.

In upsampling with a non-overlapping configuration, where the update interval $T_u$ and window duration $T_w$ are equal ($T_u = T_w$), the estimator output is interpolated to produce a denser sequence. As no new observations are introduced, the statistical content of each estimate remains unchanged, and the acceleration variance is identical to that in~\eqref{var_a}.

In post-filtering, smoothing is applied to an already-estimated sequence, which can reduce variance but simultaneously introduces additional bias and reduces effective bandwidth, degrading temporal responsiveness.

In contrast, OS-ZFM generates each estimate by solving a distinct regression problem over a shifted observation window incorporating new observations. The resulting update interval $T_u < T_w$ yields a genuinely denser estimate sequence, with acceleration variance
\begin{equation}
\mathrm{Var}(\hat{a}_{\mathrm{ovp}}) = \frac{\lambda^2\sigma_f^2}{8T_u^2},
\end{equation}

The ratio between overlapping and non-overlapping (base) configurations is thus expressed as
\begin{equation}
\frac{\mathrm{Var}(\hat{a}_{\mathrm{ovp}})}{\mathrm{Var}(\hat{a}_{\mathrm{base}})} 
= \frac{T_w^2}{T_u^2} = \frac{N^2}{M^2},
\end{equation}
which exceeds unity, reflecting the expected noise amplification from a reduced differentiation interval. However, this comparison assumes a base estimator with the same window length $N$. The key advantage of OS-ZFM lies not in variance reduction under this condition, but in its ability to decouple the noise-averaging window $N$ from the update interval $M$. For a given temporal resolution $T_u$, a non-overlapping estimator must use $T_w = T_u$, eliminating the benefit of extended averaging and yielding
\begin{equation}
\mathrm{Var}(\hat{a}_{\mathrm{base},\,T_u}) 
= \frac{\lambda^2 \sigma_f^2}{8T_u^2}.
\end{equation}

OS-ZFM instead maintains $T_w = NT > T_u = MT$, retaining the $O(N^{-3})$ variance scaling while achieving the same update rate. Consequently, OS-ZFM enables lower variance at equivalent temporal resolution by maintaining a longer observation window than any non-overlapping estimator, rather than by reducing the variance of individual estimates—a property not attainable via upsampling or post-filtering.

In practice, $N$ controls the noise--resolution trade-off, while $M$ determines the update rate. Moderate overlap improves temporal sampling while maintaining statistical performance. The induced correlation should be considered when evaluating derived quantities (e.g., velocity or acceleration), and can be mitigated by subsampling or appropriate uncertainty modeling.

To systematically evaluate the performance of OS-ZFM under controlled conditions, the signal model and simulation framework are detailed in the following section.

% =========================================================
\section{Modeled Signals}

\begin{figure}[!t]
\centering
\includegraphics[width=3.3in]{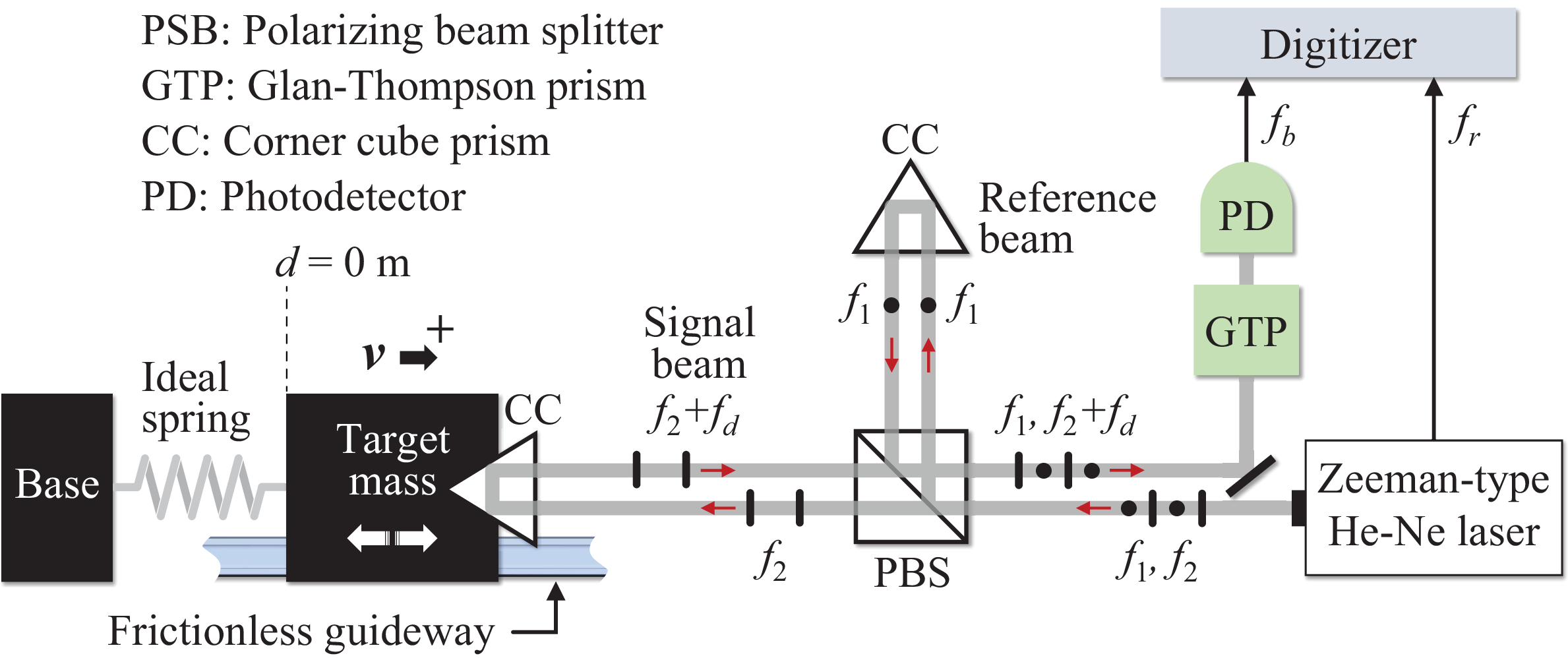}
\caption{Simulation setup of the Zeeman-type heterodyne laser Doppler interferometer. The PBS separates the orthogonally polarized frequencies $f_1$ and $f_2$ into reference and signal paths. The Doppler-shifted signal $f_d(t)$ interferes with the reference at the PD to generate the beat signal $f_b(t)$ for ZCM, HPME, MPLS, ZFM, and OS-ZFM evaluations.}
\label{sim_sys}
\end{figure}

The simulation framework is based on a Zeeman-type heterodyne interferometer, as illustrated in Fig.~\ref{sim_sys}. A target mass undergoes ideal harmonic motion, generating a Doppler-shifted beat signal. The instantaneous beat frequency is given by $f_b(t) = f_r - f_d(t)$, where $f_r = f_1 - f_2$ is the reference frequency and $f_d(t) = (2/\lambda)v(t)$ is the Doppler shift proportional to the target velocity $v(t)$.

% ---------------------------------------------------------
% ---------------------------------------------------------

The simulated measurement and reference signals, $S_b(t)$ and $S_r(t)$, incorporate intensity noise $A_r(t)$ and additive white Gaussian background noise $n(t)$ \cite{Barker1972, Carlton1981, Kowalski2001}, expressed respectively as:
\begin{equation}
\left.
\begin{aligned}
S_b(t) &= A_n(t)\sin\left(2\pi f_r t + B\sin(2\pi f_m t) + \varphi_0\right) + dc \\
S_r(t) &= A_n(t)\sin\left(2\pi f_r t + \varphi_0\right) + dc
\end{aligned}
\right\}
\label{eq:intensity}
\end{equation}
\begin{equation}
\left.
\begin{aligned}
S_b(t) &= A\sin\left(2\pi f_r t + B\sin(2\pi f_m t) + \varphi_0\right) + dc + n(t) \\
S_r(t) &= A\sin\left(2\pi f_r t + \varphi_0\right) + dc + n(t)
\end{aligned}
\right\}
\label{eq:background}
\end{equation}
where $f_m$ is the modulation frequency, $A_n(t)=A+A_r(t)$ is the instantaneous signal amplitude that combines the nominal amplitude $A$ and random white-noise fluctuation (intensity noise). $\varphi_0$ denotes the initial phase, $dc$ the direct-current offset, and $B$ the phase-modulation amplitude (rad) corresponding to the maximum Doppler shift.

The white-noise model in~\eqref{eq:background} is adopted for analytical tractability. In practical interferometric systems, colored noise, quantization, and harmonic distortion introduce correlations, non-Gaussian effects, and waveform distortions in the zero-crossing timing. As OS-ZFM performs regression over multiple zero-crossings per window, these deviations are partially averaged, capturing the dominant noise behavior. Consequently, while such non-idealities may affect absolute accuracy, the fundamental estimator behavior and performance trends remain qualitatively consistent with the idealized model.

The analytical displacement and acceleration are given by
\begin{equation}
d_s(t) = -\frac{\lambda B}{4\pi} \sin(2\pi f_m t),
\label{eq_ds}
\end{equation}
\begin{equation}
a_s(t) = \lambda\pi B f_m^2 \sin\left(2\pi f_m t\right).
\label{eq_as}
\end{equation}
The simulation parameters were specified as follows: $A=2$~V, $B=1000$~rad, $f_r=1.5$~MHz, $f_m=100$~Hz, $\varphi_0=\pi/7$, $dc=5$~V, and laser wavelength $\lambda=632.8$~nm. The signals were sampled at $f_s=20$~MHz for up to $5$~M samples, and the DC component was removed using a moving-average filter.

% ---------------------------------------------------
\begin{figure*}[!t]
\centering
\includegraphics[width=6.8in]{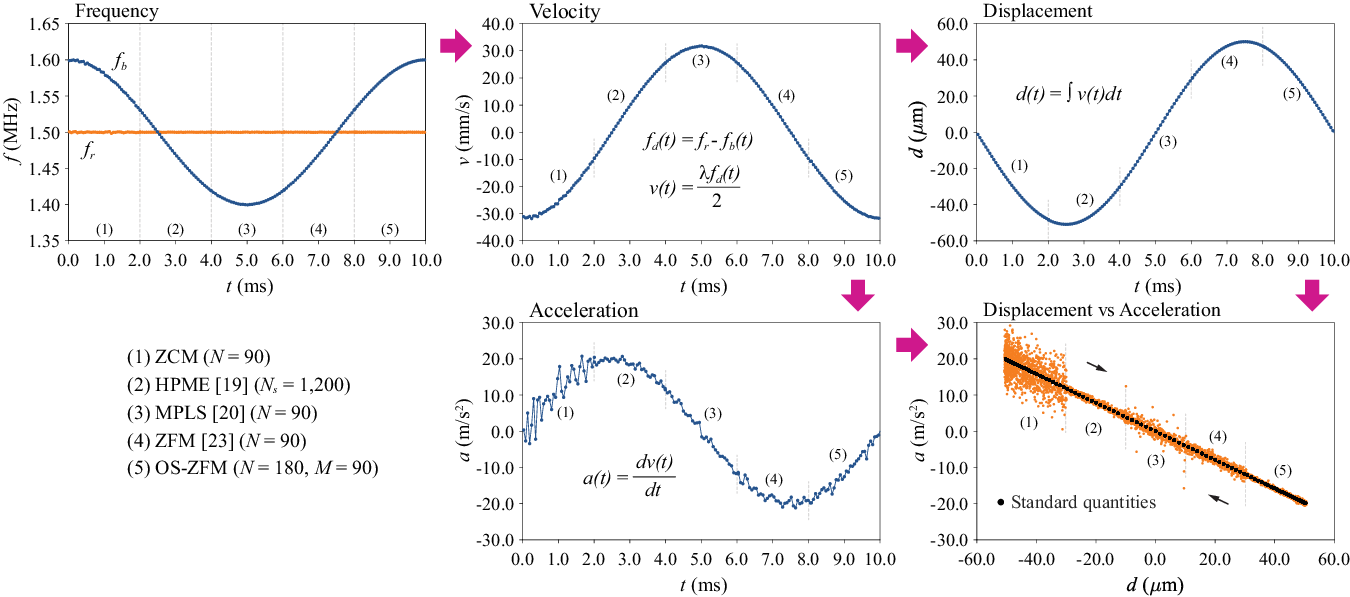}
\caption{Data processing procedure and representative estimation results under background noise (SNR = 10~dB). $f_b$ and $f_r$ are estimated using ZCM, HPME, MPLS, ZFM, and OS-ZFM, from which velocity is derived via Doppler shift, followed by integration and differentiation to obtain displacement and acceleration. The resulting acceleration--displacement trajectories are compared with analytical reference values to assess dynamic tracking accuracy.}
\label{data_pro}
\end{figure*}
% ---------------------------------------------------

The OS-ZFM observation window of $N=180$ periods ($\approx 120\ \mu$s at $f_r=1.5$~MHz) is orders of magnitude shorter than the $10$-ms modulation period, ensuring Doppler quasi-stationarity and validating the regression-based estimation. With an update interval of $M=90$ periods, this configuration balances temporal resolution and inter-estimate correlation, enabling dense frequency sampling without excessive redundancy. To ensure a rigorous comparison, ZCM was configured with $N=90$, consistent with the update interval of OS-ZFM ($M=90$). ZFM and MPLS utilized both $N=90$ and $180$ to examine window-length effects, while HPME employed corresponding window lengths ($N_s=1200$ and $2400$ samples at $f_r=1.5$~MHz and $f_s=20$~MHz), ensuring identical temporal support across all methods.

% ==============================================================
\section{Simulation Results}
\label{sim_results}

Fig.~\ref{data_pro} presents the signal processing sequence and representative tracking performance under background noise at $10$-dB SNR. Instantaneous frequencies $f_b$ and $f_r$ were estimated using the proposed OS-ZFM alongside the baseline methods. The $1.4$--$1.6$~MHz modulation of $f_b$ reflects the Doppler effect induced by harmonic motion, where $f_b = f_r$ denotes motion reversals and extrema coinciding with maximum velocity at zero displacement. From the Doppler frequency $f_d = f_r - f_b$, the velocity $v$ is derived, and subsequently integrated and differentiated to reconstruct displacement $d$ and acceleration $a$ trajectories. The resulting acceleration--displacement plot serves as a compact indicator of dynamic tracking accuracy. Analytical references and estimates are represented by black dots and scattered points, respectively.

At this low SNR level, noise-induced timing jitter causes significant dispersion in the estimated trajectories for the conventional methods. In contrast, OS-ZFM maintains close agreement with the analytical reference. This performance demonstrates that the overlapping regression framework effectively enhances statistical stability through structured noise averaging while preserving the temporal resolution required to capture rapid frequency variations.

\begin{figure*}[!t]
\centering
\includegraphics[width=6.8in]{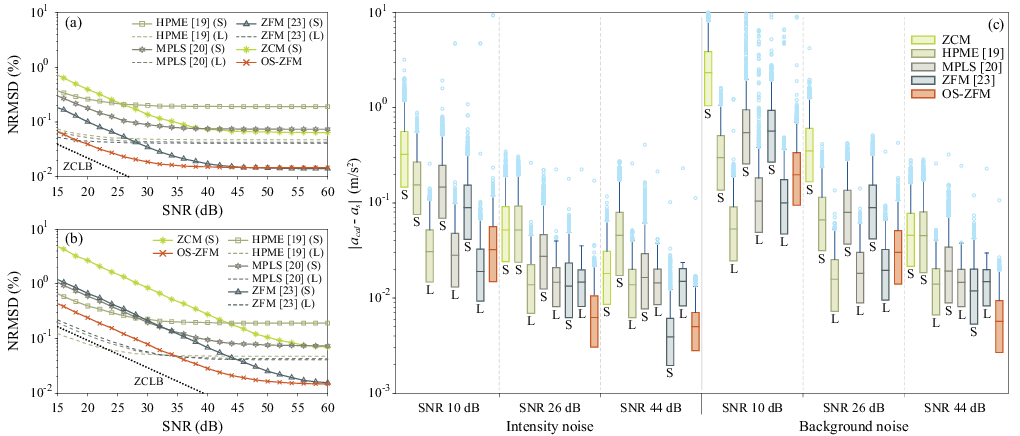}
\caption{Performance evaluation of acceleration estimation. (a)--(b) NRMSD versus SNR under intensity and background noise, respectively, where ZCLB is the theoretical zero-crossing lower bound. (c) Absolute error boxplots at SNRs of 10, 26, and 44~dB. `S' and `L' denote short ($N=90$) and long ($N=180$) window lengths for ZCM, MPLS, and ZFM, with equivalent temporal supports ($N_s=1200, 2400$) for HPME. The proposed OS-ZFM employs $N=180$ and $M=90$. Central marks, box edges, and whiskers indicate medians, interquartile ranges (IQRs), and $1.5 \times$~IQR, respectively, and markers denote outliers.}
\label{NRMSD_box}
\end{figure*}

Fig.~\ref{NRMSD_box}(a) and (b) illustrate the normalized root-mean-square deviation (NRMSD) of the estimated acceleration as a function of SNR under intensity and background noise, respectively. The NRMSD is defined as

\begin{equation}
\text{NRMSD} = \frac{\text{RMS}}{a_{s,\max} - a_{s,\min}} \times 100\%,
\label{eq:NRMSD}
\end{equation}

\noindent where $a_{s,\max}$ and $a_{s,\min}$ denote the extrema of the analytical acceleration $a_s(t)$, and RMS is the root-mean-square deviation between estimated and reference accelerations, calculated over the full time record. The maximum Doppler shift was $0.1$~MHz, corresponding to a peak acceleration of $20~\text{m/s}^2$ based on the analytical motion model.

ZCM exhibited the largest NRMSD across all SNRs due to its reliance on only two zero-crossings, making it highly sensitive to timing jitter. HPME and MPLS leverage signal structure to mitigate errors, yet they remain constrained by fixed windowing and heightened noise sensitivity during short observation intervals at low SNRs.

ZFM further improved accuracy by averaging multiple zero-crossings within each estimation window. Increasing the window length from $N = 90$ to $N = 180$ enhanced noise suppression but degraded temporal responsiveness at high SNRs.

By employing overlapping estimation windows, OS-ZFM combined the noise robustness of a long fitting window with improved temporal resolution through reduced update spacing. This structure effectively decoupled noise averaging from temporal resolution, yielding consistently lower NRMSD across all SNRs without introducing additional tracking bias.

% ------------------------------------------------------
% -------------------------------------------------------

The Zero-crossing Lower Bound (ZCLB) is included to characterize the analytical performance floor for the NRMSD of the zero-crossing-based estimations. In Fig.~\ref{NRMSD_box}(a), the ZCLB for intensity noise is significantly lower than that of background noise, reflecting the inherent suppression of multiplicative noise at the zero-crossing points, where the signal amplitude approaches zero. In Fig.~\ref{NRMSD_box}(b), under background noise, it is observed that HPME (L) slightly outperforms the ZCLB at low SNRs. This is because ZCLB represents a fundamental limit specifically for estimators relying on sparse timing information from discrete zero-crossings. In contrast, HPME exploits the dense information of the entire digitized waveform within the observation window, allowing it to access phase information beyond discrete crossing instants. Notably, the proposed OS-ZFM consistently approaches the ZCLB more closely than other zero-crossing-based methods, demonstrating its superior statistical efficiency in utilizing available timing data while maintaining high temporal resolution.

Fig.~\ref{NRMSD_box}(c) presents boxplot distributions of the absolute acceleration error $\left| a_{cal} - a_s \right|$ at SNRs of 10, 26, and 44~dB under both noise conditions. Each boxplot illustrates the error distribution for a representative realization over the full time record, where the median and interquartile range (IQR) characterize the central tendency and dispersion, respectively. At SNR = 10~dB, OS-ZFM yields significantly lower median error than all baselines for the same update interval of 90 periods. Under intensity noise, median errors are reduced from 0.089 (ZFM) and 0.146 (MPLS) to 0.032 (OS-ZFM), representing relative improvements of 64\% and 78\%, respectively. Similarly, under background noise, the error is mitigated from 0.569 (ZFM) and 0.549 (MPLS) to 0.197 (OS-ZFM)---an improvement of approximately 65\% and 64\%. Notably, OS-ZFM achieves up to a 90\% reduction relative to ZCM and consistently outperforms HPME, particularly under intensity noise.

For statistical validation, 1000 independent noise realizations were generated for each method and noise condition. A two-sided Wilcoxon signed-rank test was performed for pairwise comparisons between OS-ZFM and each baseline method at representative SNR levels. In all cases, the null hypothesis of equal median error was rejected with $p < 0.001$, indicating statistically significant improvements. Although direct overlay of theoretical bounds is not applicable due to dimensional and differentiation effects, the observed dispersion is qualitatively consistent with the uncertainty framework in Section~\ref{uncertainty}.

Increasing SNR reduces error dispersion across all methods. ZCM exhibits the widest spread due to timing jitter, while HPME and MPLS remain susceptible at low SNRs. Although ZFM improves robustness through noise averaging, OS-ZFM consistently achieves the narrowest distributions—notably under intensity noise—demonstrating superior stability through overlapping-enhanced statistical efficiency without compromising temporal resolution.

% ================================================================
\section{Experiment}
\label{exp_results}

To evaluate OS-ZFM under realistic conditions, experimental validation employs impact-induced transient motion from a mass-rubber collision. The resulting non-harmonic, high-dynamic motion introduces time-varying frequencies and nonlinear damping effects not captured in the numerical model.

\begin{figure*}[!t]
\centering
\includegraphics[width=6.8in]{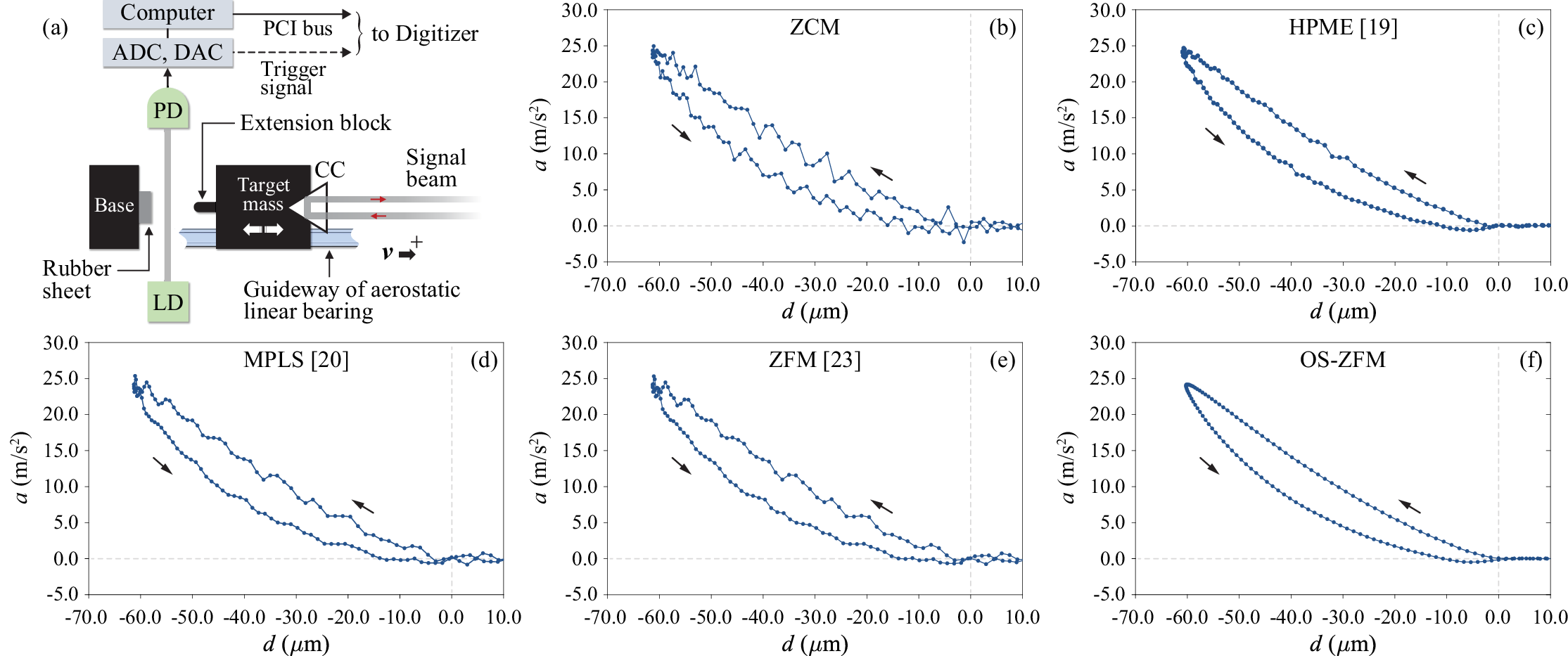}
\caption{Impact-response measurement of a target mass integrated with the moving part of an aerostatic linear bearing and a rubber sheet under test, where (a) experimental setup and (b)-(f) acceleration-displacement trajectories obtained using ZCM, HPME~\cite{Kun2025}, MPLS~\cite{Tan2024}, ZFM~\cite{Fujii2009}, and the proposed OS-ZFM, respectively. In the experiment, ZCM, MPLS, and ZFM employ $N=\text{80}$ periods, while HPME adopts $N_{s}=\text{904}$ samples to achieve an equivalent observation window. The OS-ZFM is configured with $N=\text{500}$ and $M=\text{80}$.}
\label{exp_rubber}
\end{figure*}

The impact experiment is intended as a practical transient measurement scenario to evaluate frequency estimation performance rather than collision mechanics. To ensure a direct comparison under identical conditions, all methods are applied to a single representative transient record. This approach effectively isolates the algorithmic response from physical test variability, ensuring that differences in the reconstructed trajectories stem solely from the estimator's performance. Following the configuration in Fig.~\ref{sim_sys}, the experimental setup replaces the spring-mass model with a collision-based structure as depicted in Fig.~\ref{exp_rubber}(a). A rigid mass of approximately $1.93$~kg, guided by an aerostatic linear bearing (TAAG10A-02, NTN), achieves near-frictionless motion before impacting a $3$~mm thick commercial rubber sheet. Signal detection and digitization utilize the same acquisition chain as in the numerical simulation.

Measurement is initiated by manually accelerating the target mass leftward from the right end of the aerostatic bearing, as shown in Fig.~\ref{exp_rubber}~(a). Contact between the mass extension block and the fixed rubber sheet induces a transient motion. Velocity of the mass is derived from the heterodyne Doppler signal, with acceleration and displacement subsequently obtained via numerical differentiation and integration.

% --------------------------------------------------------
% ------------------------------------------------------

Photodetector signals are digitized at $20$~MS/s with $8$--bit resolution (NI PCI-5105) for $0.25$~s, totaling $5$~M samples per channel. Data acquisition is triggered as the moving mass interrupted an optical path between a laser diode (LD) and photodiode (PD). The beat and reference frequencies are estimated using ZCM, MPLS, and ZFM with $N=80$ periods (approximately $0.045$~ms at $f_r=1.77$~MHz), while HPME utilizes an equivalent duration of $N_s=904$ samples.

The OS-ZFM parameters are set to $N=500$ and $M=80$ at $f_r=1.77$~MHz, yielding a $0.28$-ms observation window. This duration is significantly shorter than the approximately $5$~ms characteristic motion scale observed in the experiment, satisfying the quasi-stationary Doppler frequency assumption. The high zero-crossing count stabilizes the least-squares estimation through enhanced noise averaging, while $M=80$ maintains a temporal resolution comparable to the baseline methods for a rigorous evaluation.

Fig.~\ref{exp_rubber}(b)-(f) compares the acceleration-displacement trajectories reconstructed from the estimated Doppler frequency using different methods under impact-induced, high-dynamic motion. The initial impact velocity was estimated approximately $35$~mm/s. The displacement origin was defined at the impact onset, identified by a sustained acceleration increase exceeding the pre-impact noise standard deviation.

ZCM exhibited dispersion and irregularities, particularly during the impact interval, due to its reliance on discrete zero-crossing events and limited fitting. Under rapid frequency variations, acceleration estimation became sensitive to timing jitter and noise, reducing trajectory smoothness.

HPME produces smoother trajectories than ZCM by exploiting instantaneous phase information; however, residual fluctuations remain during the impact event, indicating sensitivity to amplitude modulation and noise amplification associated with the Hilbert transform under non-stationary conditions.

Both MPLS and ZFM utilize global linear fitting over a fixed observation window but differ in their local zero-crossing estimation. In this study, MPLS employs three-sample local linear regression around each zero-crossing to better accommodate rapid signal variations during impact while maintaining local linearity. Despite improving timing accuracy compared with the two-point interpolation used in ZFM, MPLS trajectories remained susceptible to noise and rapid frequency shifts.

OS-ZFM yields a coherent and physically consistent trajectory throughout the impact event. By pairing a long observation window with a short sliding update interval, it suppresses noise while preserving sensitivity to rapid dynamics. Notably, the subtle negative acceleration during the late rebound phase—reflecting the rubber's viscoelastic response—was captured with reduced fluctuation compared to baseline methods. The resulting trajectory in Fig.~\ref{exp_rubber}(f) provides a stable and smooth representation of the transient dynamics, demonstrating robust performance under practical noise conditions.

% ====================================================
\section{Discussion}

\subsection{Behavior under Transient Motion}

While the numerical simulations in Section~\ref{sim_results} demonstrate the theoretical performance of OS-ZFM under idealized harmonic conditions, the experiments in Section~\ref{exp_results} validate its behavior under practical measurement conditions. The impact-induced motion introduces non-stationary dynamics and viscoelastic effects, such as hysteresis and energy dissipation, not captured by the harmonic model. As shown in Fig.~\ref{exp_rubber}, conventional methods (ZCM, ZFM, and MPLS) exhibit noticeable fluctuations and local irregularities, particularly during acceleration sign transitions, due to sensitivity to noise and timing jitter in zero-crossing detection. HPME provides comparatively smoother trajectories; however, residual fluctuations remain during transient intervals, reflecting sensitivity to amplitude modulation and noise. In contrast, OS-ZFM yields a smoother and more stable trajectory, capturing both the rebound phase and the associated negative acceleration dip. This improvement is consistent with enhanced noise averaging from the longer regression window.

The smoothness of the OS-ZFM trajectories primarily reflects enhanced statistical stability rather than absolute physical validation. However, as the 0.28 ms observation window is significantly shorter than the $\sim$5~ms mechanical motion scale, the risk of attenuating genuine high-frequency dynamics is reduced. The stable reconstruction of the viscoelastic rebound suggests physical consistency, although quantitative verification of absolute accuracy would require an independent high-bandwidth reference measurement.

% ----------------------------------------------------------------------
\subsection{Noise Sensitivity and Statistical Uncertainty Analysis}
\label{uncertainty}

The noise sensitivity and statistical uncertainty of OS-ZFM are evaluated via numerical simulations, with the representative results. In Fig.~\ref{NRMSD_box}(c), boxplots under intensity and background noise quantify uncertainty, where the interquartile range (IQR) reflects estimation variability and outliers indicate sensitivity to noise-induced zero-crossing timing perturbations.

At SNR = 10~dB, OS-ZFM shows a narrower error distribution than methods with similar temporal support, indicating improved statistical stability under challenging noise. Uncertainty is evaluated using a Type~A approach consistent with GUM, where regression residuals estimate zero-crossing timing variance for frequency uncertainty propagation. Accordingly, the statistical uncertainty of OS-ZFM is governed by the variability of zero-crossing timing within each observation window. The $i$-th zero-crossing time in the $j$-th window is modeled as
\begin{equation}
t_i = iT_j + P_j + \epsilon_i, \qquad i = 0,1,\ldots,N,
\end{equation}
where $T_j$ and $P_j$ denote the signal period and phase offset, and $\epsilon_i$ denotes zero-mean timing error.

After least-squares fitting, the residuals $r_i = t_i - (i\hat{T}_j + \hat{P}_j)$ are used to estimate the timing variance as
\begin{equation}
\hat{\sigma}_t^2 = \frac{1}{N-1}\sum_{i=0}^{N} r_i^2,
\end{equation}
where the residual-based variance estimate $\hat{\sigma}_t^2$ provides a practical approximation of the underlying timing jitter variance $\sigma_\epsilon^2$, thereby linking the empirical uncertainty evaluation with the theoretical bias--variance analysis.

For linear regression of uniformly spaced zero-crossings, the standard uncertainty of the estimated period is given by
\begin{equation}
u(T_j) = \sqrt{\frac{12\,\hat{\sigma}_t^2}{N(N+1)(N+2)}} ,
\label{uT}
\end{equation}
and the corresponding frequency uncertainty is
\begin{equation}
u(f_j) = \frac{1}{T_j^2}
\sqrt{\frac{12\,\hat{\sigma}_t^2}{N(N+1)(N+2)}} .
\label{uf}
\end{equation}

The measured Doppler frequency $f_d(t)$ is related to the target mass velocity by $v(t)= (\lambda/2)f_d(t)$, where $\lambda$ is the optical wavelength. The corresponding acceleration is therefore obtained by
\begin{equation}
a(t)=\frac{\lambda}{2}\,\dot{f}_d(t),
\end{equation}

\noindent where $\dot{f}_d(t)$ denotes the temporal derivative of the Doppler frequency. Following GUM-based uncertainty propagation, the associated acceleration uncertainty can be expressed as
\begin{equation}
u\!\left(a(t)\right)=\frac{\lambda}{2}\,u\!\left(\dot{f}_d(t)\right),
\end{equation}
indicating its dependence on the uncertainty of the differentiated Doppler frequency. While zero-crossing timing variance determines instantaneous frequency uncertainty, numerical differentiation may amplify high-frequency noise in the estimated sequence. More specifically, the differentiation acts as a high-pass operation, amplifying high-frequency components in the estimated frequency. Consequently, acceleration uncertainty depends not only on zero-crossing timing variance but also on the spectral content of the estimate.

From Fig.~\ref{NRMSD_box}(c), the median acceleration error of OS-ZFM at SNR = 10~dB (background noise) is approximately $0.05$~m/s$^2$, corresponding to a normalized error of $2.5\times10^{-3}$ relative to the $20$~m/s$^2$ peak. Given the $1.4-1.6$~MHz range as observed in Fig.~\ref{data_pro}, this phase variability implies sub-nanosecond timing jitter. Substituting $N=180$ into \eqref{uf} yields an instantaneous frequency uncertainty of several tens of hertz, consistent with the numerical results.

% -----------------------------------------------------------
\subsection{Practical Implementation Considerations}

From an instrumentation perspective, computational efficiency and analytical transparency are as important as estimation accuracy. Unlike methods involving complex tuning or transform-domain processing, OS-ZFM applies closed-form regression directly to zero-crossing timestamps. This ensures low complexity, robustness, and predictable bias-variance behavior, essential for metrological traceability. The consistency between numerical (Fig.~\ref{NRMSD_box}) and experimental results (Fig.~\ref{exp_rubber}) demonstrates OS-ZFM as a deterministic and interpretable estimator under practical noise conditions.

OS-ZFM retains the low complexity of conventional zero-crossing estimators by employing simple algebraic updates without iterative optimization or spectral processing. In contrast to phase-based methods (e.g., HPME) that require Hilbert transforms and phase differentiation, OS-ZFM avoids transform-domain operations, reducing per-update overhead and memory demand. Although overlapping windows increase the update rate, the per-estimate computational structure remains unchanged, preserving implementation efficiency.

The computational efficiency of OS-ZFM stems from its two-stage architecture: local interpolation and global linear regression. For $N = 180$ cycles, a 50\% overlap ($M = 90$) enables the reuse of $91$ zero-crossing timestamps, requiring only $90$ new interpolations per update. Combined with a matrix-based regression using a precomputed pseudo-inverse, the total complexity is reduced to approximately $2 \times 10^3$ operations per update. At an algorithmic update rate of $16.67$~kHz (at $f_r = 1.5$~MHz), the total computational demand is approximately $3.3 \times 10^7$ operations per second. For a typical ARM Cortex-M4 class microcontroller operating at $100$~MHz (providing $\approx 125$ DMIPS), this represents a moderate processor load of approximately $33.3\%$, ensuring sufficient headroom for simultaneous tasks such as data logging or peripheral communication. %As summarized in Table~\ref{tab:complexity}, this efficiency is achieved while maintaining an NRMSD of $0.041\%$ for a representative case under $36$-dB SNR background noise, highlighting a superior accuracy--efficiency trade-off compared to the baseline methods.

Combined with its deterministic, non-iterative formulation, OS-ZFM is well suited for smart sensing and resource-constrained IoT platforms. Its low computational and energy requirements enable integration into automated calibration and distributed monitoring systems. By relying on standard interferometry without additional hardware, it reduces implementation cost while providing robust noise performance and high temporal resolution without spectral processing. Consequently, OS-ZFM serves as a low-cost, general-purpose solution for measurement and embedded sensing applications. At the system level, the interferometer is part of the baseline setup, and OS-ZFM introduces no additional hardware cost. Compared to computationally intensive alternatives and commercial estimators, it reduces processing and maintenance overhead, supporting cost-effective deployment.

While advanced time--frequency (e.g., synchrosqueezing or empirical wavelet transforms) and deep learning methods offer powerful representations, their higher computational cost and limited analytical transparency can limit real-time implementation. This work instead adopts a low-complexity, interpretable algebraic approach on zero-crossing timestamps, ensuring deterministic performance and physical interpretability for practical instrumentation.

The overlapping regression concept adopted in OS-ZFM is not restricted to zero-crossing fitting. In principle, the same framework is broadly applicable and could be extended to other time-domain methods, such as polynomial or spline-based trajectory fitting. Within these contexts, partially overlapping observation windows would similarly decouple temporal resolution from noise averaging, establishing a generalized framework for high-resolution parameter estimation in time-domain measurement systems.

% --------------------------------------------------------
\subsection{Limitations and Future Work}

While OS-ZFM improves accuracy and resolution under typical conditions, its performance involves a trade-off between noise averaging and frequency tracking. Longer windows reduce variance but may induce temporal smoothing during abrupt Doppler transitions. At sub-zero SNR levels, noise-induced timing jitter and spurious zero-crossings can degrade estimation integrity. Furthermore, the reliance on precise zero-crossing detection makes the method susceptible to waveform distortion and fringe dropouts. These limitations motivate future research into adaptive windowing to enhance tracking under non-stationary conditions and robust detection algorithms to mitigate distortion. Recent advances in adaptive windowing~\cite{Yang2023} and low-SNR zero-crossing optimization~\cite{Lian2026} offer promising frameworks for further enhancing OS-ZFM's resilience in extreme sensing environments.

% ===========================================================
\section{Conclusion}

This paper has demonstrated that OS-ZFM achieves improved frequency estimation accuracy by decoupling the noise-averaging window from the temporal update interval through structured data reuse. The proposed closed-form formulation preserves the unbiased nature and $O(N^{-3})$ variance scaling of the underlying estimator while increasing temporal sampling density. This enables consistent reduction in estimation error at equivalent temporal resolution—a capability not attainable via output upsampling or post-filtering. Experimental validation on transient impact motion demonstrates its practical effectiveness in reconstructing smooth, physically consistent acceleration--displacement profiles and revealing viscoelastic responses often obscured by noise in conventional methods. Retaining the computational efficiency of algebraic regression and avoiding iterative optimization, OS-ZFM is well suited for real-time interferometric sensing and resource-constrained embedded systems. Future work will explore adaptive windowing strategies for tracking rapidly varying or multi-component frequency signals under more demanding dynamic conditions.

\ifCLASSOPTIONcaptionsoff
  \newpage
\fi

\end{document}